\documentclass[12pt]{iopart}

\usepackage{amssymb}
\usepackage{iopams}
\usepackage{footnote}
\usepackage{hyperref}

\begin{document}

\title[]{Quantum generalized fluctuation-dissipation %
relations in terms of time-distributed observations}

\author{Yu. E. Kuzovlev}

\address{Donetsk Institute for Physics and Technology %
of NASU \\  ul. R.\,Luxemburg 72, 83114 Donetsk, Ukraine}
\ead{kuzovlev@fti.dn.ua}

\begin{abstract}
New formulations of quantum generalized fluctuation-dissipation %
relations in terms of characteristic and %
probabilistic functionals of continuous observations are %
suggested and discussed. It is shown that control %
of entropy production in quantum system turns %
any measurement in it to a source of its extra perturbations, %
and because of this effect relations between probabilities of %
mutually time-reversed processes become definitely non-local in their %
functional space.  %
\end{abstract}


\vspace{2pc} \noindent{\it Keywords}\,: %
\,generalized fluctuation-dissipation relations, %
generalized fluctuation-dissipation theorems, %
fluctuation theorems, %
quantum fluctuation-dissipation relations, %
quantum fluctuation theorems

\section{Introduction}

This paper continues recent preprint \cite{may}  %
and my earlier work \cite{fdr1} (see also %
notes \cite{z2}) devoted to quantum formulations of %
the ``generalized fluctuation-dissipation relations'' (FDR) %
\cite{bk1,bk2,bk3,ufn1} (see also references in \cite{fdr1,ufn1,z1}). %

In classical statistical mechanics (CSM) %
many different modifications of the FDR may be unified into %
one formula \cite{bkt}
\begin{eqnarray}
P(\Pi_+)\,\exp{[-\Delta S(\Pi_+)]}\, %
=\,P(\Pi_-) \,\,\,\label{mfds}
\end{eqnarray}
connecting probability $\,P(\Pi_+)\,$  %
of realization and observation of a ``forward'' process %
$\,\Pi_+\,$ with probability $\,P(\Pi_-)\,$ of its time %
reversion $\,\Pi_-\,$ and change %
$\,\Delta S(\Pi_+) =-\Delta S(\Pi_-)\,$  %
of entropy of a system under consideration %
in these processes (for details please see \cite{ufn1} %
or \cite{bkt}). %
The exponential factor here may be moved from %
left to right-hand side, - as %
$\,\exp{[-\Delta S(\Pi_-)]}$\,, - or somehow %
split between two sides,
\begin{eqnarray}
P(\Pi_+)\,\exp{[- z\,\Delta S(\Pi_+)]}\, %
=\,P(\Pi_-)\, \exp{[- (1-z)\,\Delta S(\Pi_-)]}\, %
\, \,\,\label{mfds_}
\end{eqnarray}
with some \,$z$\,, and, of course, %
this equality is quantitatively equivalent to Eq.\ref{mfds}.
In quantum statistical mechanics (QSM), however, %
such the equivalence generally is true in qualitative sense only, %
since measurements of \,$\Delta S$\, do not commute %
with that of \,$\Pi_\pm$\, and thus may influence upon %
quantities \,$P(\Pi_\pm)$\,. Our purpose below will be to show that %
this non-commutativity is actually significant at least %
for time-distributed (``continuous'') observations %
of system's evolution under external %
perturbations, and that  best (most similar) %
quantum analogue of Eqs.\ref{mfds}-\ref{mfds_} appears %
under symmetric splitting, i.e. at \,$z=1/2$\,. %
Simultaneously, our consideration will throw more light %
on physical meaning of the entropy production operator %
\,$\Delta \dot{S}(t)$\, suggested in \cite{may}. %
As the result, formally exact quantum FDR for generating - %
characteristic and probabilistic - functionals obtained in \cite{fdr1} %
now will be reformulated in more comfortable fashion  %
allowing to extend Eq.\ref{mfds_}, with \,$z=1/2$\,, %
to continuous observations in QSM.

\section{Basic quantum FDR}

In many theoretical models and physical applications %
of CSM and QSM it seems reasonable %
(or necessary \cite{bk2,ufn1}) to divide Hamiltonian %
\,$H(x)$\, (\,$x=x(t)$\,) of an externally driven system %
into two parts:
\[
H(x)\,=\,H_0 -h(x)\,\, %
\,\,\,\,\,\,\, (\,H_0=H(0)\,\,, \,\,\,\, h(0)=0\,)\,\,, \,
\]
- with \,$H_0$\, representing system's internal energy %
while \,$-h(x)$\, its interaction with sources of its %
perturbations, - and focus on changes of just the internal %
energy in their connection with entropy production in the system. %
Corresponding treatment of FDR \cite{bk1} later %
\cite{rmp,cht} was termed ``exclusive'' (on its motivations %
see e.g. \cite{ufn1} %
\footnote{\, %
In addition to historical remarks in \cite{ufn1} %
we would like to notice %
that some of scientists interpret results of their %
precursors in as broad sense as possible while another in %
maximally narrow sense. We believe to be in the first group, %
at that naturally ascribing ourselves to our precursors. %
}\,). %
For brevity, in the present %
paper we  confine ourselves just by it. Therefore we can base  %
directly on quite general quantum FDR (for Gibbs statistical %
ensembles) which were written out already in \cite{bk1} %
in Eqs.17-24. Namely, in slightly different %
but transparent notations,
\begin{eqnarray}
\langle\, A_1(t_1)\dots A_n(t_n)\, %
 e^{-H_0(t)/T} \,e^{H_0 /T}\,\rangle %
_{H_0\,,\,h(x(\tau))}\,=\,   \nonumber\\ =\, %
\langle\, \overline{A}_n(t-t_n)\dots %
\overline{A}_1(t-t_1)\, %
\, \rangle_{\overline{H}_0\,,\, %
\overline{h}(x(t-\tau))}\,\, \,\label{mr}
\end{eqnarray}
Here\, $\,A_j(t)\,$ are arbitrary operators %
in the Heisenberg representation,\, %
\begin{eqnarray}
A(t)\,=\,U^\dagger(t,0)\,A\,U(t,0) %
\,\equiv\, U^\dagger_t A\,U_t\, \,\,, \, \nonumber \\
U(t,t_0)\,=\, \overleftarrow{\exp} %
\left[ - \frac i\hbar \int_{t_0}^t %
H(x(\tau))\, d\tau\,\right]\,\,; \, \nonumber
\end{eqnarray}
the angle brackets denote average over (trace with) normalized %
canonical density matrix (statistical operator) %
$\,\rho_0 = \exp{[(F_0-H_0)/T]}\,$ %
at given ``eigen'' Hamiltonian \,$H_0$\,, %
intercation Hamiltonian \,$h(x)$\, and given %
external driving ``forces'' (Hamiltonian parameters) %
\,$x(t)$\,;\,  %
the over-line denotes transposition of operators,\, %
\,$\overline{A} = A^T\,$\, %
\footnote{\, %
Or, according to remarks on p.129 in \cite{bk1}, %
more general transform, %
  \,$\overline{A} = \Theta A^T \Theta^\dagger$\, %
with some unitary \,$\Theta$\, such that square of this %
transform is identical one, %
\,$\overline{\overline{A}} =A$\,, that is %
$\,\Theta^{\dagger T}\Theta = %
\Theta^*\Theta =1$\,. %
}\,. %
Thus, on right-hand side of Eq.\ref{mr} evolution %
of all (transposed) operators \,$\overline{A}(\tau)$\, %
is governed by Hamiltonian \,$\overline{H}(x(t-\tau))$\,. %
For brevity below without loss of generality (which is trivially %
restorable within final formulae) we assume that %
\,$\overline{H}_0 = H_0\,$ and therefore \,$H_0$\,  %
may be removed from the angle brackets' subscripts.

It should be underlined that, first,  %
arbitrary operators $\,A$\, %
in Eq.\ref{mr}, - as well as in Eqs.17-19 from \cite{bk1}, - %
may be non-Hermitian \cite{z2} %
\footnote{\, %
In particular, some of \,$A$\,'s may look like %
\,$|\mu\rangle\langle\nu|$\, with \,$|\mu\rangle$\, %
and \,$|\nu\rangle$\, being various pure quantum %
states (e.g. eigenstates of \,$H_0$\,). %
}\,. %
Second, of course, some of them may have form %
\,$\exp{(\,c\,H_0/T)}$\, with suitable numbers %
\,$c$\,. Third, clearly, Eq.\ref{mr}   %
holds true if we replace the product of \,$n$\, %
operators by arbitrary linear combination of products %
with various \,$n$\,. Exploiting this freedom %
and general properties of the trace operation,  we  %
get rights to write, instead of Eq.\ref{mr},
\begin{eqnarray}
\langle\, e^{H_0 /2T}\, {\bf B}\{V(\tau)\}\, %
 e^{-H_0(t)/T}\, {\bf A}\{V(\tau)\}\, %
e^{H_0 /2T} \,\rangle %
_{h(x(\tau))}\,=\,   \nonumber\\ =\, %
\langle\, %
\overline{{\bf B}}\{\overline{V}(t-\tau)\}\, %
 \overline{{\bf A}}\{\overline{V}(t-\tau)\}\,
\rangle_{\overline{h}(x(t-\tau))}\,\,, \,\label{mr1}
\end{eqnarray}
where \,${\bf A}\{V(\tau)\}\,$ and %
\,${\bf B}\{V(\tau)\}\,$ are some operator-valued %
expressions (functionals) composed by arbitrary collection %
of quantum variables (Heisenberg operators) \,$V(t)\,$. %
Or, making replacements
\begin{eqnarray}
{\bf A}\{V(\tau)\}\Rightarrow\, %
e^{(1-z)\,H_0(t)/2T} {\bf A}\{V(\tau)\}\, %
e^{-(1-z)\,H_0/2T}\,\,, \, \nonumber\\
{\bf B}\{V(\tau)\}\Rightarrow\, %
e^{-(1-z)\,H_0/2T} {\bf B}\{V(\tau)\}\, %
e^{(1-z)\,H_0(t)/2T}\,\,, \, \nonumber
\end{eqnarray}
formally equivalently we can write
\begin{eqnarray}
\langle\, e^{\,z\,H_0 /2T}\,
{\bf B}\{V(\tau)\}\, e^{-z\,H_0(t)/2T} %
e^{-z\,H_0(t)/2T}\, {\bf A}\{V(\tau)\}\, %
e^{\,z\,H_0 /2T} \,\rangle %
_{h(x(\tau))}\,=\,   \nonumber\\  %
=\,\langle\, %
e^{\,(1-z)\,H_0 /2T}\, %
\overline{{\bf B}}\{\overline{V}(t-\tau)\}\, %
e^{-(1-z)\, H_0(t)/2T}\, \times %
\label{mr2}\\ \times\,\, %
e^{-(1-z)\, H_0(t)/2T}\,
 \overline{{\bf A}}\{\overline{V}(t-\tau)\}\, %
e^{\,(1-z)\, H_0 /2T}\, \rangle_{\overline{h}(x(t-\tau))}\,\,
\,\nonumber
\end{eqnarray}
Evidently, Eqs.\ref{mr1}-\ref{mr2} are direct quantum analogues of %
classical generalized FDR like Eq.25 from \cite{ufn1} %
(for \,$a=b=0$\,), %
and Eq.\ref{mr2} certainly is valid at least %
at \,$0\leq \Re\,z \leq 1$\,. %

Similarly to the classical case, one can say %
that Eqs.\ref{mr1}-\ref{mr2} compare observations %
of mutually time-reversed processes conditioned %
(at least on one of two sides) by two measurements %
of system's internal energy, at beginning and end of %
observation time interval. Then difference %
of two measured quantities can be interpreted as result of %
measurement of system's internal energy change, %
\,$E(t)=H_0(t)-H_0(0)$\,, although \,$E(t)$\, is not %
quantum observable in literal sense \cite{cht}. %
Under such treatment, in view of arbitrariness of %
\,${\bf A}\{V(\tau)\}\,$ and \,${\bf B}\{V(\tau)\}\,$, %
FDR (\ref{mr1})-(\ref{mr2}) form clear ground %
for construction (definition) of quantum analogues of %
probabilistic FDR (\ref{mfds})-(\ref{mfds_}), with %
quantity \,$\Delta S$\, identified with \,$E(t)/T$\,.

\section{Time-distributed energy exchange observations    %
and their influence upon system's evolution}

Just mentioned treatment of the FDR (\ref{mr1})-(\ref{mr2}) is %
intelligent when addressed to small enough quantum %
system (or, to be precise, statistical ensembles of small %
systems) but seems rather doubtful in respect to large ones, %
in particular, consisting of a small subsystem in contact with  %
thermal bath (thermostat). If external perturbations are applied %
to the subsystem then observation of its behavior is  %
at once (indirect) continuous observation of energy flow %
through it from external ``work sources'' %
to thermostat, even though %
the latter eventually may accept all the work.   %
Therefore it would be more adequate approach to practice %
if we reformulated FDR (\ref{mr1})-(\ref{mr2}) in terms of %
time-local ``internal energy change rate'' (IECR) \cite{may} %
and related entropy production.

This task all the more is actual since in quantum theory  %
changes of system's entropy may be determined sooner by a number %
of accepted (dissipated) quanta than their summary energy, %
so that in general the change \,$E(t)=H_0(t)-H_0(0)$\, hardly %
determines \,$\Delta S$\,. This circumstance is visually %
reflected by  formulae (22) and (24) from \cite{may}, %
as well as by formula (27) from \cite{fdr1}. %

Taking in mind that time-distributed measurements %
of \,$\Delta S$\, naturally are ordered in time, %
we have to resort to a kind of chronological %
ordering of quantum variables (operators) under %
statistical averaging.
For certainty, we may choose the %
well grounded ``Jordan-symmetrized chronological'' %
operator ordering rule (see e.g. \cite{may,fdr1,z2} and %
references in \cite{fdr1}). Then the functionals %
\,${\bf A}\{V(\tau)\}\,$ and \,${\bf B}\{V(\tau)\}$\, %
can be chosen e.g. in the form
\begin{eqnarray}
{\bf A}\{V(\tau)\}\,=\, %
\overleftarrow{\exp}\,[\, %
\frac 12 \int_0^t a(\tau)\cdot %
V(\tau)\,d\tau\,]\,\,, \,\nonumber\\  %
{\bf B}\{V(\tau)\}\,=\, %
\overrightarrow{\exp}\,[\, %
\frac 12 \int_0^t a(\tau)\cdot %
V(\tau)\,d\tau\,]\,\,, \,\label{ab}  %
\end{eqnarray}
where \,$V(t)=U^\dagger_t V \,U_t$\, %
is some collection of variables (operators),\,  %
\,$a(t)$\, are {\it c}-number valued %
(generally complex) probe functions, and \,$\cdot$\, %
means summation (``scalar product'' or ``convolution'' %
of two arrays). %

First, consider left and right-hand sides of Eq.\ref{mr2} %
after substitution of expressions (\ref{ab}). %
On the left, let us make transformations as follows,  %
\begin{eqnarray}
\langle\, e^{\,z\, H_0 /2T}\,
{\bf B}\{V(\tau)\}\, %
 e^{-z\,H_0(t)/T}\, {\bf A}\{V(\tau)\}\, %
e^{\,z\,H_0 /2T} \,\rangle %
_{h(x(\tau))}\,=\,   \label{ls}\\ =\,
\langle\, e^{\,z\, H_0 /2T}\,
\overrightarrow{\exp}\,\{\, %
 \int_0^t [\,\frac i\hbar\, H(x(\tau)) + \frac {a(\tau)}2\cdot %
V\,]\,d\tau\,\}\, %
e^{-z\,H_0/2T}\, \, \times\, \, \nonumber\\
\times\, e^{-z\,H_0/2T}\, %
\overleftarrow{\exp}\,\{\, %
 \int_0^t [\,-\frac i\hbar\,H(x(\tau)) + \frac {a(\tau)}2\cdot %
V\,]\,d\tau\,\}\, %
e^{\,z\,H_0 /2T} \,\rangle %
_{h(x(\tau))}\,=\,   \nonumber\\
=\, \langle\, \overrightarrow{\exp}\,\{\, %
 \int_0^t [\,\frac i\hbar\, e_z\,H(x(\tau))\,e_z^{-1} + %
\frac {a(\tau)}2\cdot %
e_z V\,e_z^{-1}\,]\,d\tau\,\}\, %
\, \times\, \, \nonumber\\
\times\,\, \overleftarrow{\exp}\,\{\, %
 \int_0^t [\,-\frac i\hbar\, e_z^{-1}\,H(x(\tau))\,e_z + %
\frac {a(\tau)}2\cdot %
e_z^{-1} V\,e_z\,]\,d\tau\,\}\, %
\rangle_{h(x(\tau))}\,\,, \,\nonumber
\end{eqnarray}
with\, \,$e_z\equiv e^{\,z\,H_0 /2T}$\, %
and \,$e_z^{-1}\equiv e^{-z\,H_0 /2T}$\,. %

Quite similarly, but with \,$e_{1-z}$\, in place %
of \,$e_{z}$\,, transforms right side of Eq.\ref{mr2}. %

Second, notice that instead of Hermitian operator %
\,$H(x(t))$\, in Eq.\ref{ls} two generally non-Hermitian %
operators have appeared, \,$e_z^{-1}\,H(x(\tau))\,e_z$\,  %
and its conjugation %
\footnote{\, %
Here and below we neglect differences %
between concepts of Hermitian, self-conjugated, %
self-adjoint, etc., operators and everywhere exploit %
mere term ``Hermitian''.}\,.
To feel meaning of such transformations, - %
which already were in use in \cite{fdr1}, - %
let us set there \,$a=0$\, and confine our %
present consideration by case of {\it \,real\,} %
\,$z$\,. Then, after simple %
algebraic manipulations, expression (\ref{ls}) turns to
\begin{eqnarray}
\langle\, e^{\,z\, H_0 /2T}\,
e^{-z\,H_0(t)/T}\,e^{\,z\,H_0 /2T} \,\rangle %
_{h(x(\tau))}\,=\,  \label{ls0}\\
=\, %
\langle\, \overrightarrow{\exp}\,\{\, %
 \int_0^t [\,\frac i\hbar\, H(z,x(\tau)) - %
\frac 12\,G(z,x(\tau))\,]\,d\tau\,\}\, %
\times\, \nonumber\\ \times\, \,
\overleftarrow{\exp}\,\{\, %
 \int_0^t [\,-\frac i\hbar\, H(z,x(\tau)) - %
\frac 12\,G(z,x(\tau))\,]\,d\tau\,\}\, %
\rangle_{h(x(\tau))}\,\,, \,\nonumber
\end{eqnarray}
with operators \,$H(z,x)$\, and \,$G(z,x)$\, %
defined by
\begin{eqnarray}
H(z,x)\,=\,H_0-h(z,x)\,\,, \,\,\,\, \nonumber\\ %
h(z,x)\,=\, \frac 12\,[\,e_z h(x)\,e_z^{-1} + %
e_z^{-1} h(x)\,e_z\,]\,\,, \,\label{hz}\\
G(z,x)\,=\, %
\frac i\hbar\,[\,e_z h(x)\,e_z^{-1} - %
e_z^{-1} h(x)\,e_z\,]\,\, \,\label{gz}
\end{eqnarray}
Thus, \,$H(z,x)=H_0-h(z,x)$\, and \,$G(z,x)$\, %
both are Hermitian operators formed by Hermitian and %
anti-Hermitian components of \,$e_z h(x)\,e_z^{-1}$\, %
respectively.
It will be comfortable to characterize such operators  %
by their matrix elements in basis formed by %
eigenstates of \,$H_0$\,. Then Eqs.\ref{hz}-\ref{gz} %
read
\begin{eqnarray}
h_{\mu\nu}(z,x)\,=\, h_{\mu\nu}(x)\, \cosh\, %
\frac {z E_{\mu\nu}}{2T}\,\,, \,\,\nonumber\\  %
G_{\mu\nu}(z,x)\,=\, %
\frac {2i}\hbar\, h_{\mu\nu}(x)\, \sinh\, %
\frac {z E_{\mu\nu}}{2T}\,\,, \,\label{hg}
\end{eqnarray}
with \,$E_{\mu\nu}=E_{\mu} -E_{\nu}$\, and %
\,$E_{\mu}$\, being \,$H_0$\,'s eigenvalues: %
\,$H_0|\mu\rangle =E_\mu |\mu\rangle$\,.

Eq.\ref{ls0} evidently prompts to rewrite it   %
in the spirit of Jordan-symmetrized chronological %
ordering rule, as
\begin{eqnarray}
\langle\, e^{\,z\, H_0 /2T}\,
e^{-z\,H_0(t)/T}\,e^{\,z\,H_0 /2T} \,\rangle %
_{h(x(\tau))}\,=\,  \label{lsj}\\
=\, %
\texttt{Tr}\,\,\, \overleftarrow{\exp}\,\{\, %
 \int_0^t [\,-\frac i\hbar\, H(z,x(\tau)) - %
\frac 12\,G(z,x(\tau))\,]\,d\tau\,\}\,\, %
\times\, \nonumber\\ \times\,\, %
\rho_0\,\, %
\overrightarrow{\exp}\,\{\, %
 \int_0^t [\,\frac i\hbar\, H(z,x(\tau)) - %
\frac 12\,G(z,x(\tau))\,]\,d\tau\,\}\, %
=\, \nonumber\\ =\, %
\langle\, \exp\,[\, %
 -\int_0^t U_\tau^\dagger  %
G(z,x(\tau))\,U_\tau\,d\tau\,]\, %
\rangle_{\,h(z,x(\tau))}\,\,, \,\nonumber
\end{eqnarray}
with \,$U_t^\dagger G(z,x(t))\,U_t$\,  playing role %
of continuously measured quantum variable (observable) %
and unitary evolution operator \,$U_t$\, %
now being determined by modified (effective) %
Hamiltonian \,$H(z,x)=H_0-h(z,x)$\,. %

Before writing out result of analogous transformation %
of right-hand side of Eq.\ref{mr2}, %
for brevity and visuality %
let us assume, as usually, that the external forces %
possess definite parities \,$\epsilon =\pm 1$\, in respect %
to time reversal, that is
\begin{eqnarray}
\overline{h}(x)\,=\, h(\epsilon x)\,\, \,\label{par}
\end{eqnarray}
Then at \,$a(\tau)=0$\,, %
i.e. at \,${\bf A}\{\cdot\}={\bf B}\{\cdot\}=1$\,, %
on the right in  Eq.\ref{mr2} we have

\begin{eqnarray}
\langle\, e^{(1-z)\, H_0 /2T}\,
e^{-(1-z)\,H_0(t)/T}\,e^{(1-z)\,H_0 /2T} \,\rangle %
_{h(\epsilon x(t-\tau))}\,=\,  \label{rsj}\\
=\, %
\langle\, \exp\,[\, %
 -\int_0^t U_\tau^\dagger  %
G(1-z,\epsilon x(t-\tau))\,U_\tau\,d\tau\,]\, %
\rangle_{\,h(1-z,\epsilon x(t-\tau))}\,\,, \,\nonumber
\end{eqnarray}
now with evolution governed by effective %
Hamiltonian \,$H(1-z,\epsilon x)$\,.

Hence, the particular case of Eq.\ref{mr2},
\begin{eqnarray}
\langle\, e^{\,z\, H_0 /2T}\,
e^{-z\,H_0(t)/T}\,e^{\,z\,H_0 /2T} \,\rangle %
_{h(x(\tau))}\,=\,  \label{pc}\\ =\,
\langle\, e^{(1-z)\, H_0 /2T}\,
e^{-(1-z)\,H_0(t)/T}\,e^{(1-z)\,H_0 /2T} \,\rangle %
_{h(\epsilon x(t-\tau))}\,\,, \,\nonumber
\end{eqnarray}
can be expressed, - for real \,$z$\,, - also by equality
\begin{eqnarray}
\langle\, \exp\,[\, %
 -\int_0^t U_\tau^\dagger  %
G(z,x(\tau))\,U_\tau\,d\tau\,]\, %
\rangle_{\,h(z,x(\tau))}\, %
\,=\,  \label{pcj}\\ =\, %
\langle\, \exp\,[\, %
 -\int_0^t U_\tau^\dagger  %
G(1-z,\epsilon x(t-\tau))\,U_\tau\,d\tau\,]\, %
\rangle_{\,h(1-z,\epsilon x(t-\tau))}\, \,\nonumber
\end{eqnarray}
This equality is formally identical to Eq.\ref{pc} %
but, in contrast to it, describes continuous observations of %
system's energy exchange with sources of external perturbations. %
At that, variables \,$G(z,x)$\, and \,$G(1-z,x)$\, delegate %
the IEC's rate (IECR) but, clearly, not in literal sense. %
Naive direct interpretation of \,$G(z,x)$\, %
is possible in the classical %
limit only when, according to Eqs.\ref{hz}-\ref{gz}, %
\begin{eqnarray}
h(z,x)\,\rightarrow\,h(x)\,\,, \,\,\,\,\, %
G(z,x)\,\rightarrow\,\frac zT\,\mathcal{L}_0 \,h(x)\,=\, %
z\,\Delta\dot{S}(x)\,\,, \,\nonumber\\ %
\Delta\dot{S}(x)\,=\,\frac 1T\, \frac i\hbar %
\,[H_0,h(x)]\,=\, \frac 1T\, \frac i\hbar\, %
[H(x),H_0]\, %
\,, \,\nonumber
\end{eqnarray}
with\, \,$\Delta\dot{S}(x)$\, %
having sense of entropy production per unit time, %
or simply entropy production (EP), and %
\,$\mathcal{L}_0$\, being unperturbed Liouville %
super-operator defined by
\[
\mathcal{L}_0\, A\,=\, \frac i\hbar\,[H_0,A]\,\,, \,\,\,\,\, %
(\mathcal{L}_0\, A)_{\mu\nu}\,=\, %
\frac i\hbar\,E_{\mu\nu}\,A_{\mu\nu}\,\,
\]
In essentially quantum situations, however, %
\,$h(z,x)\neq h(x)$\, and \,$G(z,x)$\, is not %
proportional to \,$z$\,, which means that in fact %
EP depends on conditions, or ``intensity'', of  %
its measurements.

Moreover, relation (\ref{pcj}), together with (\ref{hz})-(\ref{gz}), %
shows that in general its left and right sides concern,  %
strictly speaking, two different systems with different %
interaction Hamiltonians. But practical control of external %
parameters of interaction Hamiltonian %
(IH) usually in no way means control %
of its detail structure (all matrix elements). %
Therefore it seems reasonable to restrict further consideration %
by special choices of the ``observation parameter'' \,$z$\,. %
Clearly, such are first of all \,$z=1$\, and \,$z=1/2$\,.

At \,$z=1$\,, Eq.\ref{pcj} reduces to
\begin{eqnarray}
\langle\, \exp\,[\, %
 -\int_0^t U_\tau^\dagger  %
G(1,x(\tau))\,U_\tau\,d\tau\,]\, %
\rangle_{\,h(1,x(\tau))}\, %
\,=\,1\,\,  \label{pcj1} %
\end{eqnarray}
From viewpoint of this relation in itself, %
the effective, - ``renormalized by observation'', - %
IH \,$h(1,x)$\, is nothing but system's actual IH. %
Therefore, making redesignation %
\,$h(1,x) \Rightarrow h(x)$\,, we can rewrite %
Eq.\ref{pcj1} in the form
\begin{eqnarray}
\langle\, \exp\,[\, %
 -\int_0^t U_\tau^\dagger  %
\Delta\dot{S}(x(\tau))\,U_\tau\,d\tau\,]\, %
\rangle_{\,h(x(\tau))}\, %
\,=\,1\,\,, \,  \label{qbk} %
\end{eqnarray}
where EP \,$\Delta\dot{S}(x)$\, is defined by  %
\begin{eqnarray}
\Delta\dot{S}_{\mu\nu}(x)\,\equiv\, %
G_{\mu\nu}(1,x)\,=\, %
\frac {2i}\hbar\, h_{\mu\nu}(1,x)\, \tanh\, %
\frac {E_{\mu\nu}}{2T}\, \Rightarrow\,  %
\nonumber\\ \Rightarrow\,  %
\frac {2i}\hbar\, h_{\mu\nu}(x)\, \tanh\, %
\frac {E_{\mu\nu}}{2T}\, \,\label{ep1}
\end{eqnarray}
This (above mentioned) statistical equality %
was derived originally in \cite{fdr1} and  %
in different way recently in \cite{may}. %

\section{Time-symmetric observations and entropy %
production operator}

At \,$z=1/2$\,, in framework of analogous treatment of  %
renormalized IH \,$h(1/2,x)$\, 
as factual IH, i.e. after redesignation %
\,$h(1/2,x)\Rightarrow h(x)$\,, %
Eq.\ref{pcj} yields
\begin{eqnarray}
\langle\, \exp\,[\, %
 - \int_0^t U_\tau^\dagger\,\frac 12\,   %
\Delta\dot{S}(x(\tau))\,U_\tau\,d\tau\,]\, %
\rangle_{\,h(x(\tau))}\, %
\,=\,  \label{pcj12}\\ =\, %
\langle\, \exp\,[\, %
 - \int_0^t U_\tau^\dagger\,\frac 12 \,  %
\Delta\dot{S}(\epsilon x(t-\tau))\,U_\tau\,d\tau\,]\, %
\rangle_{\,h(\epsilon x(t-\tau))}\, \,, \,\nonumber
\end{eqnarray}
where now by definition
\begin{eqnarray}
\frac 12\, \Delta\dot{S}_{\mu\nu}(x)\,\equiv\, %
G_{\mu\nu}(1/2,x(t))\,=\, %
\frac {2i}\hbar\, h_{\mu\nu}(1/2,x)\, \tanh\, %
\frac {E_{\mu\nu}}{4T}\, \Rightarrow\,  %
\nonumber\\ \Rightarrow\, %
\frac {2i}\hbar\, \,h_{\mu\nu}(x)\, \tanh\, %
\frac {E_{\mu\nu}}{4T}\, \,\label{ep12} %
\end{eqnarray}
This is quantum version of classical equality %
following from FDR (\ref{mfds_}) after %
its integration over all possible processes (observations).

Notice that Eq.\ref{par} together with Eq.\ref{ep12} %
or \ref{ep1} implies anti-symmetry property
\begin{eqnarray}
\Delta\dot{S}(\epsilon x)\,=\, %
-\,\overline{\Delta\dot{S}}(x)\,\,, \,\,\,\,\, %
\Delta\dot{S}_{\mu\nu}(\epsilon x)\,=\, %
-\Delta\dot{S}_{\nu\mu}(x)\,\,, \label{as}
\end{eqnarray}
ensuring that changes of system's entropy %
in mutually time-reversed processes %
(under time-reversed external conditions) %
differ by their signs only.

Now, let us consider at \,$z=1/2$\, %
simultaneously both sides of FDR (\ref{mr2}) with
functionals \,${\bf B}\{V(\tau)\}\,$ and %
 \,${\bf A}\{V(\tau)\}$\, taken in the form (\ref{ab}). %
It reads
\begin{eqnarray}
\langle\, \overrightarrow{\exp}\,\{\, %
 \int_0^t [\,\frac i\hbar\, e\,H(x(\tau))\,e^{-1} + %
\frac {a(\tau)}2\cdot %
e\, V\,e^{-1}\,]\,d\tau\,\}\, %
\, \times\, \, \nonumber\\
\times\,\, \overleftarrow{\exp}\,\{\, %
 \int_0^t [\,-\frac i\hbar\, e^{-1}\,H(x(\tau))\,e + %
\frac {a(\tau)}2\cdot %
e^{-1} V\,e\,]\,d\tau\,\}\, %
\rangle\,=\, \, \label{lr0}\\
=\, %
\langle\, \overrightarrow{\exp}\,\{\, %
 \int_0^t [\,\frac i\hbar\, e\,H(\epsilon x(t-\tau))\,e^{-1} + %
\frac {a(t-\tau)}2\cdot %
e\, \overline{V}\,e^{-1}\,]\,d\tau\,\}\, %
\, \times\, \, \nonumber\\
\times\,\, \overleftarrow{\exp}\,\{\, %
 \int_0^t [\,-\frac i\hbar\, e^{-1}\,H(\epsilon x(t-\tau))\,e + %
\frac {a(t-\tau)}2\cdot %
e^{-1} \overline{V}\,e\,]\,d\tau\,\}\, %
\rangle\,\,, \,\nonumber
\end{eqnarray}
where\, \,$e\equiv e_{1/2}= \exp{(H_0/4T)}$\, and %
\,$e^{-1}\equiv e_{1/2}^{-1}= \exp{(-H_0/4T)}$\,, we used our %
conventions (\ref{par}) and %
\,$\overline{H}_0=H_0$\,, and omitted the angle %
brackets' subscripts because they are %
superfluous here. %

Then perform at Eq.\ref{lr0} transformations %
like (\ref{hz})-(\ref{gz}), %
assuming that variables under consideration possess %
definite parities,
\[
\overline{V}\,=\,\varepsilon\,V\,\,\,\,\, %
(\,\varepsilon\pm 1\,)\,\,, \,
\]
and introducing short notations
\begin{eqnarray}
\widetilde{x}(\tau)\,=\,\epsilon\,x(t-\tau)\,\,, %
\,\,\,\,\, \widetilde{a}(\tau)\,=\, %
\varepsilon\,a(t-\tau)\,\,,\, \label{tilde}\\
V_+ \,=\,\frac 12\,[\,e^{-1} V\,e\,+\, %
e\, V\,e^{-1}\,]\,\,, \, \label{v}\\ %
V_- \,=\,\frac 12\,[\,e^{-1} V\,e\, - %
e\, V\,e^{-1}\,]\,\, \, %
\,\,\,(\,e\equiv e^{H_0/4T}\,)\,\, \nonumber
\end{eqnarray}
After that Eq.\ref{lr0} turns to
\begin{eqnarray}
\langle\, \overrightarrow{\exp}\,\{\, %
 \int_0^t [\,\frac i\hbar\, H^\prime(x(\tau),a(\tau)) - %
\frac 12\, G(1/2,x(\tau)) + %
\frac {a(\tau)}2\cdot V_+\,]\,d\tau\,\}\, %
\, \times\, \, \nonumber\\
\times\,\, \overleftarrow{\exp}\,\{\, %
 \int_0^t [\,-\frac i\hbar\, H^\prime(x(\tau),a(\tau)) - %
\frac 12\, G(1/2,x(\tau)) + %
\frac {a(\tau)}2\cdot V_+\,]\,d\tau\,\}\, %
\rangle\,=\, \, \nonumber \\
=\, %
\langle\, \overrightarrow{\exp}\,\{\, %
 \int_0^t [\,\frac i\hbar\, %
H^\prime(\widetilde{x}(\tau),\widetilde{a}(\tau)) - %
\frac 12\, G(1/2,\widetilde{x}(\tau)) + %
\frac {\widetilde{a}(\tau)}2\cdot V_+\,]\,d\tau\,\}\, %
\, \times\, \, \nonumber\\ \times\,\, %
\overleftarrow{\exp}\,\{\, %
 \int_0^t [\,-\,\frac i\hbar\, %
H^\prime(\widetilde{x}(\tau),\widetilde{a}(\tau)) - %
\frac 12\, G(1/2,\widetilde{x}(\tau)) + %
\frac {\widetilde{a}(\tau)}2\cdot V_+\,]\,d\tau\,\}\, %
\rangle\,\, \,\label{lr1}
\end{eqnarray}
withe effective ``renormalized'' Hamiltonian %
\,$H^\prime$\, is expressed by formulae
\begin{eqnarray}
H^\prime(x,a)\,=\,H_0- h^\prime(x,a)\,\,, \,\,\, \nonumber\\ %
h^\prime(x,a)\,=\, h_+(x)\, + %
\frac \hbar{2i}\,a\cdot V_-\,\,, \,\label{h12}\\
h_+(x)\,=\, \frac 12\,[\,e\, h(x)\,e^{-1} + %
e^{-1}h(x)\,e\,]\,\,, \, \nonumber\\
h_{\mu\nu}^\prime(x,a)\,=\, h_{\mu\nu}(x)\, \cosh\, %
\frac {E_{\mu\nu}}{4T}\, + \, %
\frac {i\hbar}2\, a\cdot V_{\mu\nu} \,\sinh\,  %
\frac {E_{\mu\nu}}{4T}\,\, \,\label{hmn}
\end{eqnarray}
Thus,  naturally, continuous observations of arbitrary variables %
\,$V$\, also influence upon system's behavior, - %
by changing its effective IH, 
\,$h(x)\Rightarrow h^\prime(x,a)$\,, - %
so that probe functions \,$a(t)$\, acquire role %
of additional external forces. %

It is important to underline %
that left and right-hand effective IHs in Eq.\ref{mr1} %
are not mutually transposed, i.e. %
\,$h^\prime(\epsilon x,\varepsilon a)\neq %
\overline{h^\prime}(x,a)$\,. In fact, %
\,$\overline{h^\prime}(x,a)= %
h^\prime(\epsilon x,-\varepsilon a)$\,, %
because \,$V_-$\,'s parity is opposite to that of \,$V$\,. %
In this sense, observations (as well as perturbations) %
violate time symmetry of observed processes, %
even in spite of symmetry due to choice\,$z=1/2$\,. %

\section{Time symmetrized %
quantum characteristic and probabilistic functionals}

In order to reformulate Eq.\ref{lr1} in more plausible and standard form, %
it is reasonable to make ``back renormalization'' of %
both the IH and observed variables, by replacements %
like already used above,
\begin{eqnarray}
h_+(x)\,\Rightarrow\,h(x)\,\,, \,\,\,\,
V_+\,\Rightarrow\,V\,\,, \nonumber\\
h_{\mu\nu}(x)\, \cosh\, %
\frac {E_{\mu\nu}}{4T}\,\Rightarrow\, %
h_{\mu\nu}(x)\,\,, \,\,\,\,\, %
V_{\mu\nu}\, \cosh\, %
\frac {E_{\mu\nu}}{4T}\,\Rightarrow\, %
V_{\mu\nu}\,\,, \,\nonumber
\end{eqnarray}
Besides, it is comfortable to introduce %
super-operator \,$\mathcal{T}$\, defined by
\begin{eqnarray}
(\,\mathcal{T}A\,)_{\mu\nu}\,=\, %
A_{\mu\nu}\, \tanh\, %
\frac {E_{\mu\nu}}{4T}\,\equiv\, %
\mathcal{T}_{\mu\nu}\,A_{\mu\nu}\,\, \,\label{t}
\end{eqnarray}
After that effective Hamiltonian becomes
\begin{eqnarray}
H^\prime(x,a)\,\Rightarrow\, %
H(x,a)=H_0-h(x,a)\,\,, \,\,\,\,\, %
h^\prime(x,a)\,\Rightarrow\,h(x,a)\,\,, \, \nonumber\\
h(x,a)\,\equiv\, h(x)\, + %
\frac {i\hbar}{2}\,a\cdot \mathcal{T}\,V\,\,, \, \label{hp}\\
h_{\mu\nu}(x,a)\,=\, h_{\mu\nu}(x)\, %
+ \, \frac {i\hbar}2\, a\cdot V_{\mu\nu} \,\tanh\,  %
\frac {E_{\mu\nu}}{4T}\,\,, \,\label{hpmn}
\end{eqnarray}
and Eq.\ref{lr1} can be rewritten as
\begin{eqnarray}
\langle\, \overrightarrow{\exp}\,\{\, %
 \int_0^t [\,\frac i\hbar\, H(x(\tau),a(\tau)) - %
\frac 14\, \Delta\dot{S}(x(\tau)) + %
\frac {a(\tau)}2\cdot V\,]\,d\tau\,\}\, %
\, \times\, \, \nonumber\\
\times\,\, \overleftarrow{\exp}\,\{\, %
 \int_0^t [\,-\frac i\hbar\, H(x(\tau),a(\tau)) - %
\frac 14\, \Delta\dot{S}(x(\tau)) + %
\frac {a(\tau)}2\cdot V\,]\,d\tau\,\}\, %
\rangle\,=\, \, \nonumber\\
=\, %
\langle\, \overrightarrow{\exp}\,\{\, %
 \int_0^t [\,\frac i\hbar\, %
H(\widetilde{x}(\tau),\widetilde{a}(\tau)) - %
\frac 14\, \Delta\dot{S}(\widetilde{x}(\tau)) + %
\frac {\widetilde{a}(\tau)}2\cdot V\,]\,d\tau\,\}\, %
\, \times\, \, \nonumber\\ \times\,\, %
\overleftarrow{\exp}\,\{\, %
 \int_0^t [\,-\,\frac i\hbar\, %
H(\widetilde{x}(\tau),\widetilde{a}(\tau)) - %
\frac 14\, \Delta\dot{S}(\widetilde{x}(\tau)) + %
\frac {\widetilde{a}(\tau)}2\cdot V\,]\,d\tau\,\}\, %
\rangle\,\,, \,\label{lr2}
\end{eqnarray}
with operator \,$\Delta\dot{S}(x)/2$\, already introduced %
in (\ref{ep12}):  %
\begin{eqnarray}
\frac 12\, \Delta\dot{S}(x)\,=\, %
\frac {2i}\hbar\, \mathcal{T}\,h(x)\,\, \,\label{de} %
\end{eqnarray}


Further, for more visuality and  %
without loss of generality (see e.g. remarks in \cite{ufn1}), %
let us choose the ``seed'' IH \,$h(x)$\, in ``bilinear'' form
\begin{eqnarray}
h(x)\,=\,x\cdot Q\,\,, \,\label{lh}
\end{eqnarray}
and introduce variables
\begin{eqnarray}
I=\mathcal{L}Q\,\,, \,\,\,\,\, %
J=\mathcal{L}V\,\, \,\,\,\,\, %
\left(\,\mathcal{L}\equiv\,  %
\frac {4iT}\hbar\, \mathcal{T} =\, %
\frac {4iT}\hbar\,\tanh\, %
\frac {\hbar\mathcal{L}_0}{4iT} \right)\, \,\label{ij}
\end{eqnarray}
Then the ``half of EP'' operator (\ref{de}) and %
effective IH \,$h(x,a)$\, determined by %
(\ref{hp}) become
\begin{eqnarray}
\frac 12\, \Delta\dot{S}(x)\,=\, %
\frac 1{2T}\,x\cdot I\,\,, \,\label{de_}\\  %
h(x,a)\,=\, x\cdot Q\, +\, %
\frac {\hbar^2}{8T}\, a\cdot J\,\, \, \label{eh0}
\end{eqnarray}

Taking in mind conventions of the Jordan-symmetrized %
chronological operator ordering rule, instead of Eq.\ref{lr2} %
we can write also simply
\begin{eqnarray}
\langle\, \exp\,\{\, %
 \int_0^t [\,- \frac 1{2T}\, %
x(\tau)\cdot I(\tau)\, + %
a(\tau)\cdot V(\tau)\,]\,d\tau\,\}\, %
\rangle_{\,h(x(\tau),\,a(\tau))}\,=\, %
\, \label{scf}\\ =\, %
\langle\, \exp\,\{\, %
 \int_0^t [\,- \frac 1{2T}\, %
\widetilde{x}(\tau)\cdot I(\tau)\, + %
\widetilde{a}(\tau)\cdot V(\tau)\,]\,d\tau\,\}\, %
\rangle_{\,h(\widetilde{x}(\tau),\,\widetilde{a}(\tau))}\, %
\,, \,\nonumber
\end{eqnarray}
where \,$I(\tau)$\, and \,$V(\tau)$\, %
are meant as commutative (\,{\it c}-number valued) %
random processes imaging \,{\it Heisenberg\,} operator variables %
\,$U_\tau^\dagger I\,U_\tau$\, %
and \,$U_\tau^\dagger V\,U_\tau$\,, %
with evolution operator \,$U_\tau$\, concretized %
by angle brackets' subscripts. %
It is easy to extend FDR (\ref{lr1})-(\ref{scf}) %
to operator variables \,$V$\, which are time %
dependent already in Schr\"{o}dinger %
representation, so that then time argument of  %
classical image variable \,$V(t)$\, in %
Eq.\ref{scf} delegates complete ``double'' time %
dependency of its quantum originals. %

The characteristic functional FDR (\ref{scf}) %
in essence is symmetrized version of relation (20)-(21)  %
from \cite{fdr1}. These relations show that under %
time-distributed observations the perturbation %
parameters (external forces) and observation parameters %
(probe functions) inevitably get entangled and %
partially swop their roles.
In contrast to the classical limit, the %
entangling is mutual and not exact since it is accompanied by %
``deformation'' of operator variables due to factors implied %
by the super-operator \,$\mathcal{T}$\, (\ref{t}). %
In particular, clearly, variables \,$I$\, and \,$J$\, (\ref{ij}) %
are nothing but (unperturbed) time derivatives of  %
\,$Q$\, and \,$V$\,, respectively, deformed by smoothing %
over time intervals \,$\sim \hbar/T$\,. %

Notice, besides, that at imaginary probe functions, %
\,$a(t)=i\xi(t)$\,, or complex ones the effective IH %
\,$h(x,a)$\, becomes non-Hermitian. But this one more unpleasant %
peculiarity of quantum case in principle %
does not prevent transformation of FDR (\ref{scf}) for %
characteristic functionals (CF) into FDR for %
probabilistic functionals (PF). %

With such purpose, introduce %
auxiliary effective IH  %
\begin{eqnarray}
h^\prime(x,y)\,\equiv\, x\cdot Q\, + %
\,y\cdot J\,\,, \, \label{eh}
\end{eqnarray}
with \,$J$\, defined in (\ref{ij}), %
so that
\[
h(x,a)\,=\,h^\prime\left(x, %
\,\frac {\hbar^2}{8T}\,a\right)\,\,, \,
\]
and functional of four (sets of) variables as follows,
\begin{eqnarray}
\Xi\{a(\tau),b(\tau)\,|\,x(\tau),y(\tau)\}\,=\, %
\label{cf}\\ =\, %
\langle\, \exp\,\{\, %
 \int_0^t [\,a(\tau)\cdot V(\tau)\,+\, %
b(\tau)\cdot I(\tau)\,]\,d\tau\,\}\, %
\rangle_{\,h^\prime(x(\tau),\,y(\tau))}\,\, \, \nonumber
\end{eqnarray}
Evidently, this is joint CF of variables \,$V$\, %
and \,$I$\, evolving under perturbations described by %
effective IH \,$h^\prime(x,y)$\, with arbitrary additional forces %
\,$y(t)$\, in place of forces \,$(\hbar^2/8T)\,a(t)$\, %
from (\ref{eh0}) related to probe functions. %
In terms of new CF (\ref{cf}), FDR (\ref{scf}) reads
\begin{eqnarray}
\Xi\{a,-\frac x{2T}\,|\,x,\frac {\hbar^2}{8T}\,a\}\,=\, %
\Xi\{\widetilde{a},-\frac {\widetilde{x}}{2T}\,|\, %
\widetilde{x},\frac {\hbar^2}{8T}\,\widetilde{a}\}\,
\,, \, \label{scf_}
\end{eqnarray}
thus reducing CF (\ref{cf}) of four %
independent (collections of) arguments %
to functional of twice lesser number of arguments %
and entangling perturbations and observations. %

Then, consider joint PF of classical %
images of \,$V(t)$\, and \,$I(t)$\,,
\begin{eqnarray}
P\{V(\tau),I(\tau)\,|\,x(\tau),y(\tau)\}\,=\, \label{pf}\\  %
=\, \int_\xi \int_\eta \exp{\{\,-\int [\,i\xi(\tau)\cdot V(\tau) + %
i\eta(\tau)\cdot I(\tau)\,]\,d\tau\,\}}\, %
\times \,\nonumber\\ \,\times\,\, %
\Xi\{i\xi(\tau),i\eta(\tau)\,|\,x(\tau),y(\tau)\}\, %
\,, \,  \nonumber
\end{eqnarray}
where \,$\int_\xi\int_\eta \dots\,$\, means functional Fourier %
transform. In terms of this PF, for left side %
of Eq.\ref{scf_} we have
\begin{eqnarray}
\Xi\{i\xi(\tau),-\frac {x(\tau)}{2T}\,|\, %
x(\tau),\,\frac {\hbar^2}{8T}\, i\xi(\tau)\}\,=\, %
\label{lpf}\\ =\, %
\int _V \int_I \exp{\{\,\int [\,i\xi\cdot V - %
\frac 1{2T}\,x\cdot I\,]\,d\tau\,\}}\, %
P\{V,I\,|\,x,\,\frac {\hbar^2}{8T}\,i\xi\}\, %
\,, \, \nonumber
\end{eqnarray}
where \,$\int_V \dots\,$\, represents inverse %
Fourier transform over \,$V$\,. Inverting it back, %
after simple formal manipulations we obtain %
\begin{eqnarray}
\int_\xi \exp{[\,-\int i\xi\cdot V\,d\tau\,]}\, %
\,\Xi\{i\xi,-\frac x{2T}\,|\, %
x,\,\frac {\hbar^2}{8T}\,i\xi\}\,  =\, \label{lpf_}\\ =\, %
\int_\xi \exp{\{\,\int [\,-i\xi\cdot V %
+ \, \frac {\hbar^2}{8T}\, %
i\xi\cdot \frac {\delta}{\delta y}\,]\, %
d\tau\,\}}\,\, %
\times \nonumber\\ \times \,\, %
\Xi\{i\xi,-\frac x{2T}\,|\,x,y\}\, %
\,\|_{y=0}  \,\,=\, \nonumber\\ =\, %
\int_I \exp{[-\frac 1{2T}\int x\cdot I\,d\tau\,]}\, %
\,P\{V -\frac {\hbar^2}{8T}\, \frac {\delta}{\delta y}\, %
,I\,|\,x,y\}\, \,\|_{y=0} \,\, %
\,=\, \nonumber\\ =\, %
\exp{[\,-\int d\tau\, \frac {\hbar^2}{8T}\, %
\frac {\delta}{\delta V} %
\cdot \frac {\delta}{\delta y}\, %
]}\, 
\int_I \exp{[-\frac 1{2T}\int x\cdot I\,d\tau\,]}\, %
\,P\{V,I\,|\,x,y\}\, \,\|_{y=0} \, \nonumber
\end{eqnarray}
Quite similar manipulations can be made with %
right-hand side of Eq.\ref{scf_}  %
resulting in
\begin{eqnarray}
\int_I \exp{[-\frac 1{2T}\int x\cdot I\,d\tau\,]}\, %
\,P\{V -\frac {\hbar^2}{8T}\, \frac {\delta}{\delta y}\, %
,I\,|\,x,y\}\, \,\|_{y=0} \,\, %
\,=\, \nonumber\\ =\, %
\int_I \exp{[-\frac 1{2T}\int \widetilde{x}\cdot I\,d\tau\,]}\, %
\,P\{\widetilde{V} -\frac {\hbar^2}{8T}\, %
\frac {\delta}{\delta \widetilde{y}}\, %
,I\,|\,\widetilde{x},\widetilde{y}\}\, %
\,\|_{\widetilde{y}=0} \,\, \,\label{pfr}
\end{eqnarray}
with \,$\widetilde{V}(\tau)=\varepsilon V(t-\tau)$\, %
and \,$\widetilde{y}(\tau)=\varepsilon y(t-\tau)$\,. %
This is symmetrized analogue of relations (30)-(31) from \cite{fdr1}.

The coefficient \,$\,\hbar^2/8T$\, in these relations %
is  the only factor evidently including the Planck constant %
and therefore determines most significant differences %
from classical limit. %
Under the latter, \,$I\rightarrow \mathcal{L}_0 Q$\,, %
\,$J\rightarrow \mathcal{L}_0 V$\,, and Eq.\ref{pfr} %
reduces to
\begin{eqnarray}
\int_I \exp{[-\frac 1{2T}\int x\cdot I\,d\tau\,]}\, %
\,P\{V,I\,|\,x,0\}\, =\,  \nonumber\\ =\, %
\int_I \exp{[-\frac 1{2T}\int \widetilde{x}\cdot I\,d\tau\,]}\, %
\,P\{\widetilde{V},I\,|\,\widetilde{x},0\}\, %
\, \,\label{cpfr}
\end{eqnarray}
thus canceling the auxiliary forces \,$y$\,. %

Probabilistic quantum FDR (\ref{pfr}) strongly simplifies %
under choice \,$V\Rightarrow I=\mathcal{L}Q= %
(4iT/\hbar)\mathcal{T}Q$\,, %
when \,$V$\, become (time-smoothed) time derivatives %
of the variables  \,$Q$\, conjugated with actual ``seed'' %
external forces \,$x$\,. In this interesting special %
case, clearly, \,$\varepsilon =-\epsilon$\,, %
and Eq.\ref{scf_} shortens to relation
\begin{eqnarray}
\Xi\{a-\frac x{2T}\,|\,x,\frac {\hbar^2}{8T}\,a\}\,=\, %
\Xi\{\widetilde{a}-\frac {\widetilde{x}}{2T}\,|\, %
\widetilde{x},\frac {\hbar^2}{8T}\,\widetilde{a}\}\,
\, \, \label{sscf}
\end{eqnarray}
for functional of three arguments,

\begin{eqnarray}
\Xi\{a(\tau)\,|\,x(\tau),y(\tau)\}\,=\, %
\label{scf}\\ =\, %
\langle\, \exp\,[\, %
 \int_0^t a(\tau)\cdot I(\tau)\,d\tau\,]\, %
\rangle_{\,h^\prime(x(\tau),\,y(\tau))}\,\,, \, \nonumber
\end{eqnarray}
with \,$I=\mathcal{L}Q=(4iT/\hbar)\,\mathcal{T}Q$\,, %
\,$\widetilde{a}(\tau)=-\epsilon a(t-\tau)$\,, %
and effective IH
\[
h^\prime(x,y)\,=\,x\cdot Q\,+\, %
y\cdot \mathcal{L}^2\, Q\,\,
\]
Introducing PF \,$P\{I\,|\,x,y\}$\, generated %
by CF (\ref{scf}) and repeating above manipulations, %
one can come to relation
\begin{eqnarray}
\exp{[\,-\int d\tau\, \frac {\hbar^2}{8T}\, %
\frac {\delta}{\delta I} %
\cdot \frac {\delta}{\delta y}\, %
]}\,\, %
P_0\{I\,|\,x,y\}\, \,\|_{y=0} \,\, %
\,=\,  \nonumber\\ =\, %
\exp{[\,-\int d\tau\, \frac {\hbar^2}{8T}\, %
\frac {\delta}{\delta \widetilde{I}} %
\cdot \frac {\delta}{\delta \widetilde{y}}\, %
]}\,\, %
P_0\{\widetilde{I}\,\,|\,\widetilde{x},\widetilde{y}\}\, %
\,\|_{\widetilde{y}=0} \,\,, \,\label{spfr}
\end{eqnarray}
where
\begin{eqnarray}
P_0\{I\,|\,x,y\}\,\equiv\, P\{I\,|\,x,y\}\, %
\exp{[-\frac 1{2T}\int x\cdot I\,d\tau\,]}\, %
\, \,\label{p0}
\end{eqnarray}
and\, \,$\widetilde{I}(\tau)= %
-\epsilon I(t-\tau)$\,,\, %
\,$\widetilde{y}(\tau)= -\epsilon y(t-\tau)$\,. %

This is quantum generalization of classical FDR %
for probabilistic functional of ``currents'' %
(or ``flows``, or ``velocities'') \,$I(t)=dQ(t)/dt$\, %
conjugated with external forces, which for the first time %
were derived in \cite{bk1,bk2}. %
It is easy to generalize relations %
(\ref{sscf}) and (\ref{spfr})-(\ref{p0}) %
to cases when along with \,$I$\, some other variables %
\,$V$\, are under observations as in CF (\ref{cf}).  %

In comparison wuth classical case, Eq.\ref{spfr} looks unclosed %
since involves virtual excess arguments, - i.e. forces \,$y(t)$\,, - %
which eventually are frozen at zero. But we can  %
make Eq.\ref{spfr} formally closed if sufficiently %
expand collection of quantum variables under observation. %
For instance, if take
\begin{eqnarray}
Q\,=\,\{Q^+_{\mu\nu},Q^-_{\mu\nu}\}\,\,, \,\,\nonumber\\ %
Q^+_{\mu\nu}= q_{\mu\nu}\,\frac {|\mu\rangle\langle \nu| + %
|\nu\rangle\langle \mu|}{\sqrt{2}}\,\,, \,\,\,\,\, %
Q^-_{\mu\nu}= q_{\mu\nu}\, \frac {|\mu\rangle\langle \nu| - %
|\nu\rangle\langle \mu|}{i\sqrt{2}}\,\,, \,\nonumber
\end{eqnarray}
with some real symmetric \,$q_{\mu\nu}\,$, %
and, respectively, \,$x=\{x^+_{\mu\nu},x^-_{\mu\nu}\}$\, and %
\begin{eqnarray}
I\,=\,\{I^+_{\mu\nu},I^-_{\mu\nu}\}\,=\, %
\frac {4iT}\hbar\,\mathcal{T}_{\mu\nu}\, %
\{-Q^-_{\mu\nu},Q^+_{\mu\nu}\}\,\,, \, %
\,\nonumber\\
J\,=\,\{J^+_{\mu\nu},J^-_{\mu\nu}\}\,=\, %
- \left( \frac {4T}\hbar\,\mathcal{T}_{\mu\nu} \right)^2\, %
\{Q^+_{\mu\nu},Q^-_{\mu\nu}\}\,\, \, \nonumber %
\end{eqnarray}
Clearly, now
\[
\frac {\delta}{\delta y_{\mu\nu}} \,=\, %
- \left( \frac {4T}\hbar\,\mathcal{T}_{\mu\nu} \right)^2\, %
\frac \delta{\delta x_{\mu\nu}}\,\,, \, %
\]
and Eq.\ref{spfr} simplifies to
\begin{eqnarray}
\exp{[\, 2T\int d\tau\, \sum\, %
\frac {\delta}{\delta I^\pm_{\mu\nu}}\, %
\mathcal{T}^2_{\mu\nu}\, \frac {\delta}{\delta x^\pm_{\mu\nu}}\, %
]}\,\, %
P_0\{I\,|\,x\}\,=\,  \nonumber\\ =\, %
\exp{[\,2T\int d\tau\, \sum\, %
\frac {\delta}{\delta \widetilde{I}^\pm_{\mu\nu}}\, %
\mathcal{T}^2_{\mu\nu}\, %
\frac {\delta}{\delta \widetilde{x}^\pm_{\mu\nu}}\, %
]}\,\, %
P_0\{\widetilde{I}\,\,|\,\widetilde{x}\}\, %
\,, \,\label{gpfr}
\end{eqnarray}
where\, \,$\widetilde{I}^\pm_{\mu\nu}(\tau)= %
\mp I^\pm_{\mu\nu}(t-\tau)$\,,\, %
\,$\widetilde{x}^\pm_{\mu\nu}(\tau)= \pm %
x^\pm_{\mu\nu}(t-\tau)$\,, and, as before,  %
\begin{eqnarray}
P_0\{I\,|\,x\}\,\equiv\, P\{I\,|\,x\}\, %
\exp{[-\frac 1{2T}\int \sum\, %
x^\pm_{\mu\nu}\, I^\pm_{\mu\nu}\,d\tau\,]}\, %
\, \,\label{gp0}
\end{eqnarray}
Notice that all the variables \,$Q$\,, \,$I$\,, %
\,$J$\, have finite classical limits at %
\,$\hbar\rightarrow 0$\,, therefore role of %
common characteristic parameter of their %
``quantumness'' now is played by factor %
\,$\mathcal{T}^2_{\mu\nu}\,$ in Eq.\ref{gpfr}.

These relations possibly may be better visualized %
with the help of easy derivable operator equality
\begin{eqnarray}
e^{A\nabla_X\nabla_Y}\,e^{B\,XY}\,=\, %
\frac 1{1-AB}\,\times\,
\label{oe}\\ \times \, %
\left\{\,\exp{\left[\, %
\frac {B\,XY}{1-AB} + \frac {AB}{1-AB}\,  %
(X\nabla_X + Y\nabla_Y) + %
\frac {A\,\nabla_X \nabla_Y}{1-AB}\, %
\right]}\,\right\}\,\,, \,\nonumber
\end{eqnarray}
where the braces mean normal ordering of operators %
(all differentiation, or ``annihilation'', operators are on the %
right from the multiplication, or ``creation'', ones). %
Correspondingly,
\begin{eqnarray}
e^{A\nabla_X\nabla_Y}\,e^{B\,XY}\,P(X,Y)\,=\, %
\frac 1{1-AB}\, %
\exp{\left(\frac {B\,XY}{1-AB}\right)}\, %
\times \,\label{fe}\\ \times\,\, %
\exp{[\,A\,(1-AB)\,\nabla_X \nabla_Y\,]}\, %
P\left(\frac X{1-AB},\frac Y{1-AB}\right)\,\, \nonumber
\end{eqnarray}
with formally arbitrary function \,$P(X,Y)$\,. %
Applying these formulae to Eqs.\ref{gpfr}-\ref{gp0}, %
we can write, symbolically but quite transparently,
\begin{eqnarray}
\exp{[\,-\frac 1{2T}\int \frac {x\,I} %
{1+\mathcal{T}^2}\,d\tau\,]}\, %
\times \,\nonumber\\ \times\,\, %
\exp{[\,2T \int d\tau\, %
\frac {\delta}{\delta I}\, %
\mathcal{T}^2\,(1+\mathcal{T}^2)\, %
\frac {\delta}{\delta x}\, %
]}\,\, %
P\left(\frac I{1+\mathcal{T}^2}\,|\, %
\frac x{1+\mathcal{T}^2}\right)\, %
 =\, \nonumber\\ =\, %
\exp{[\,-\frac 1{2T}\int \frac %
{\widetilde{x}\,\widetilde{I}} %
{1+\mathcal{T}^2}\,d\tau\,]}\, %
\times \,\label{gr}\\ \times\,\, %
\exp{[\,2T \int d\tau\, %
\frac {\delta}{\delta \widetilde{I}}\, %
\mathcal{T}^2\,(1+\mathcal{T}^2)\, %
\frac {\delta}{\delta \widetilde{x}}\, %
]}\,\, %
P\left(\frac {\widetilde{I}}{1+\mathcal{T}^2}\,|\, %
\frac {\widetilde{x}}{1+\mathcal{T}^2}\right)\, \, \,\nonumber
\end{eqnarray}
Other, more physically meaningful, %
representations of the probabilistic FDR %
(\ref{gpfr}) = (\ref{gr}) will be considered separately.

In terms of CF, FDR (\ref{gpfr}) = (\ref{gr}) %
become much more nicely: Eq.\ref{sscf} takes form
\begin{eqnarray}
\Xi\{a-\frac x{2T}\,|\,x -2T\mathcal{T}^2 a\}\,=\, %
\Xi\{\widetilde{a}-\frac {\widetilde{x}}{2T}\,|\, %
\widetilde{x} -2T\mathcal{T}^2\widetilde{a}\}\,
\,, \, \label{gcf}
\end{eqnarray}
where parities of \,$a$\,'s are opposite to that %
of \,$x$\,'s. If we are interested in FDR for various %
many-time statistical moments (or cumulants) and %
response functions of \,$I$\,'s and related variables, %
the Eq.\ref{gcf} may be good start point.

\section{Discussion and resume}

Our above results show that analysis of differences %
between quantum FDR %
for continuous observations and similar classical %
FDR can be reduced to analysis of specific symmetry relations %
like Eq.\ref{spfr}, - involving virtual auxiliary forces %
\,$y(t)$\,, - in comparison with their %
classical limit at \,$\hbar^2 \rightarrow 0$\,, for instance,
\begin{eqnarray}
P_0\{I\,|\,x,0\}\,=\, P_0\{\widetilde{I}\,|\,%
\widetilde{x},0\}\, \, \,\label{cr}
\end{eqnarray}
Indeed, according to Eq.\ref{p0}, in both quantum and %
classical theories we have
\begin{eqnarray}
\frac {P\{I\,|\,x\}}{P\{\widetilde{I}\,|\,\widetilde{x},0\}}\, %
=\, %
\frac {P_0\{I\,|\,x\}} %
{P_0\{\widetilde{I}\,|\,\widetilde{x}\}}\, %
\exp{[\frac 1{T}\int x\cdot I\,d\tau\,]}\, %
\,, \,\label{qr}
\end{eqnarray}
where in quantum case \,$P\{I\,|\,x\}\equiv P\{I\,|\,x,y=0\}$\, %
and \,$P_0\{I\,|\,x\}\equiv P_0\{I\,|\,x,y=0\}$\,, %
and we took into account that
\[
\int_0^t \widetilde{x}\cdot \widetilde{I}\,d\tau\, %
=\, -\int_0^t x\cdot I\,d\tau\,
\]
because of opposite parities of the forces and currents. %
In classical case, evidently, %
right-hand fraction in Eq.\ref{qr} is identically unit, %
\,$P_0\{I\,|\,x\}/P_0\{\widetilde{I}\,|\,\widetilde{x}\} %
\Rightarrow 1$\,. %
In quantum theory this equality, that is relation (\ref{cr}), %
seemingly does not contradict Eq.\ref{spfr} but at the same %
time does not follow from it. What is for the Eq.\ref{gpfr},  %
it is certainly inconsistent with(\ref{cr}).

Thus, in the framework of %
continuous quantum observations there is no general  %
``one-to-one'' correspondence between probabilities of %
mutually time-reversed processes. Instead of simple %
{\it \,c}\,-number exponential proportionality %
coefficient, they are connected through definite %
integral operators acting in functional space of %
observations and perturbations. %

This is not surprising, for anybody knows that in quantum %
mechanics any measurements causes %
more or less unpredictable excess perturbation of observed system. %
Therefore, naturally, comparison of observations of mutually %
time-reversed processes can not be as literal %
as in classical theory.  %
It requires control %
of total change in system's %
(internal) energy or entropy. But if such %
the control is included into consideration then the artificial %
excess perturbation manifests itself evidently as %
effective renormalization of system's Hamiltonian. %
This results in mutual entanglement of perturbations %
and observations and thus in the mentioned complication %
of probabilistic FDR.

To resume, %
we expounded several different forms of quantum %
generalized fluctuation-dissipation relations (FDR) %
which express symmetry of laws of quantum dynamics %
in respect to time reversal. Then we focused on %
formulation of FDR for continuously time-distributed %
observations of energy exchange %
and entropy production rates and other %
variables in externally driven systems. %

It is demonstrated that one and the same statistical %
ensemble of quantum processes requires different %
Hamiltonian and entropy production operators to be %
described in terms of discrete measurements or continuous %
measurements, because of their influence %
upon evolution of quantum system. %
Naturally, this effect %
is significant at high frequencies \,$\gtrsim T/\hbar$\, only, %
and just they determine peculiarities of quantum FDR in %
comparison with the classical ones.

In particular, operator \,$\Delta\dot{S}$\, %
of entropy production (EP) per unit time differs from operator %
of external work per unit time (divided by temperature \,$T$\,) %
by suppressing contributions of high-frequency quantum %
transitions. This result coincides with one obtained in %
another way in \cite{may} and means merely that %
at high frequencies the EP operator counts number of quanta %
rather than their energy.

Maximum visual similarity of quantum FDR for %
continuous observations to their classical prototypes is %
achieved when mutually and time-reversed processes %
symmetrically involve EP measurements and therefore %
equally renormalize system's Hamiltonian. In such %
representation, differences of quantum from classical FDR %
reduce to definite entanglement of observations and %
perturbations at high frequencies, which is %
expressed by rather simple formulae.  %

Nevertheless, the obtained probabilistic quantum FDR still %
are not quite comfortable for heuristic or analytical %
exploitation and need in more careful mathematical %
investigations.

%

\,\,\,

---------------------------

\,\,\,


\begin{thebibliography}{11}

\bibitem{may}
Kuzovlev Yu.E.\, arXiv\,  1305.3533

\bibitem{fdr1}
Kuzovlev Yu.E.\, arXiv\,  cond-mat/0501630

\bibitem{z2} Kuzovlev Yu.E.\,  arXiv\, 1108.1740

\bibitem{bk1}
Bochkov G.N. and Kuzovlev Yu.E. %
Sov.Phys.-JETP {\bf 45}\, 125 (1977) %

\url{http://www.jetp.ac.ru/cgi-bin/dn/e_045_01_0125.pdf}

\bibitem{bk2}
Bochkov G.N. and Kuzovlev Yu.E.\, Sov.Phys.-JETP {\bf 49}\, 543 (1979) %

\url{http://www.jetp.ac.ru/cgi-bin/dn/e_049_03_0543.pdf}

\bibitem{bk3}
Bochkov G.N. and Kuzovlev Yu.E.\, Physica {\bf A106} 443 (1981)
[Preprint NIRFI 138 (Gorkii USSR 1980)]

\bibitem{ufn1}
Bochkov G N and Kuzovlev Yu E,\, %
Phys. Usp. {\bf 56} (6) (2013); DOI: 10.3367/UFNe.0183.201306d.0617

arXiv\, 1208.1202

\bibitem{z1} Kuzovlev Yu.E.\,  arXiv\,  1106.0589

\bibitem{bkt}
Bochkov G.N.\, Kuzovlev Yu.E. and Troitskii V.S.\, %
Soviet Physics Doklady {\bf 29} 458 (1984) %
[DAN SSSR {\bf 276} 854 (1984)]


\bibitem{rmp}
Esposito M.\, Harbola U.\, and Mukamel S.\,
Rev. Mod. Phys. {\bf 81} 1665 (2009)

\bibitem{cht}
Campisi M.\, H\"{a}nggi P.\, and Talkner P.\,
Rev. Mod. Phys. {\bf 83} 771 (2011)




\end{thebibliography}
\end{document}